\documentclass[fleqn,twoside]{article}
\usepackage[headings]{espcrc2}
\usepackage{graphicx}
\title{Comment on the paper "Search for oscillation of the electron-capture decay probability of $^{142}$Pm" at {\it arXiv:0807.0649v1}}
\author{
Yu.A.~Litvinov\address[GSI]{Gesellschaft f\"ur Schwerionenforschung GSI, 64291 Darmstadt, Germany}%
\thanks{E-mail: y.litvinov@gsi.de}, %
F.~Bosch\addressmark[GSI], %
N.~Winckler\addressmark[GSI]\address[JLU]{Justus-Liebig Universit{\"a}t, 35392 Gie{\ss}en, Germany}, %
D.~Boutin\addressmark[JLU], %
H.G.~Essel\addressmark[GSI], %
T.~Faestermann\address[TUM]{Technische Universit\"at M\"unchen, 85748 Garching, Germany}, %
H.~Geissel\addressmark[GSI]\addressmark[JLU], %
S.~Hess\addressmark[GSI], %
P.~Kienle\addressmark[TUM]\address[SMI]{Stefan Meyer Institut f\"ur subatomare Physik, 1090 Vienna, Austria}, %
R.~Kn{\"o}bel\addressmark[GSI]\addressmark[JLU], %
C.~Kozhuharov\addressmark[GSI], %
J.~Kurcewicz\addressmark[GSI], %
L.~Maier\addressmark[TUM], %
K.~Beckert\addressmark[GSI], %
C.~Brandau\addressmark[GSI], %
L.~Chen\addressmark[JLU], %
C.~Dimopoulou\addressmark[GSI], %
B.~Fabian\addressmark[JLU], %
A.~Fragner\addressmark[SMI], %
E.~Haettner\addressmark[JLU], %
M.~Hausmann\address[MSU]{Michigan State University, East Lansing, Mi 48824, U.S.A.}, %
S.A.~Litvinov\addressmark[GSI]\addressmark[JLU], %
M.~Mazzocco\addressmark[GSI]\address[Pad]{Dipartimento di Fisica, INFN, I35131, Padova, Italy}, %
F.~Montes\addressmark[MSU], %
A.~Musumarra\address[CAT]{INFN-Laboratori Nazionali del Sud, I95123 Catania, Italy}%
\address[CAT2]{Universit\'a di Catania, I95123 Catania, Italy}, %
C.~Nociforo\addressmark[GSI], %
F.~Nolden\addressmark[GSI], %
W.R.~Pla{\ss}\addressmark[GSI]\addressmark[JLU], %
A.~Prochazka\addressmark[GSI], %
R.~Reda\addressmark[SMI], %
R.~Reuschl\addressmark[GSI], %
C.~Scheidenberger\addressmark[GSI]\addressmark[JLU], %
M.~Steck\addressmark[GSI], %
T.~St\"ohlker\addressmark[GSI]\address[Hei]{Ruprecht-Karls Universit{\"a}t Heidelberg, 69120 Heidelberg, Germany}, %
S.~Torilov\address[SPB]{St. Petersburg State University, 198504 St. Petersburg, Russia}, %
M.~Trassinelli\addressmark[GSI]\address[PAR1]{INP, CNRS UMR 7588, France and Universit{\'e} Pierre at Marie Curie-Paris6, Paris, France}, %
B.~Sun\addressmark[GSI]%
\address[PEK]{Peking University, Beijing 100871, China}, %
H.~Weick\addressmark[GSI], %
M.~Winkler\addressmark[GSI]}%
\runtitle{}%
\runauthor{}

\begin{document}

\begin{abstract}
It is argued that orbital electron-capture decays of neutral
$^{142}$Pm atoms implanted into the lattice of a solid (LBNL
experiment) do not fulfil the constraints of {\it true two-body
beta decays}, since momentum as well as energy of the final state
are distributed among three objects, namely the electron neutrino,
the recoiling daughter atom and {\it the lattice phonons}. To our
understanding, this could be a reason for the non-observation of a
periodic time modulation in the number of electron-capture decays
of implanted neutral $^{142}$Pm atoms.
\vspace{1pc}%
\end{abstract}

\maketitle
\par%
The authors report on a measurement \cite{LBNL} at the 88 inch
LBNL cyclotron of the orbital electron-capture (EC) decay
probability of $^{142}$Pm atoms implanted into a metallic matrix
(most probably in the neutral charge state immediately after
implantation).  They found a pure exponential decrease of the
number of EC decays per time unit, without a significant periodic
modulation of the decay curve. This is seemingly in disagreement
to our findings at GSI \cite{GSI} for the EC decay of stored and
cooled hydrogen-like $^{142}$Pm$^{58+}$ ions, where a time
modulation with a period of 7 seconds and a (normalized) amplitude
of about 0.2 was observed. In searching for possible reasons of
the diverging results, the authors discuss for instance, whether
the remaining electrons--in the Berkeley case--"could provide a
decoherence of the neutrino momentum states in the larger phase
space of the final atomic states after the decay". It is argued
that this is most probably not the case, since "our experiment
detected K-shell x-rays, meaning that the captured electron was
indeed a K-shell electron with a similar wavefunction to the
hydrogenic ions investigated" [at GSI].
\par%
We have no objections against this reasoning. Moreover, we
appreciate the very carefully planned and conducted experiment at
Berkeley as well as its detailed description and interpretation.
However, we want to emphasize what is in our opinion the
fundamental difference of an experiment observing EC decays of
implanted neutral atoms on the one hand, and an experiment
recording EC decays of 'free' hydrogen-like ions (albeit confined
by magnetic fields) on the other hand. In the former case we have
not a true two-body decay, since in the final state momentum as
well as energy are distributed among three objects, the electron
neutrino, the recoiling daughter atom (recoil energy of about 90
eV for the daughter atom $^{142}$Nd) {\it and the lattice
phonons}. The recoil energy of the daughter atom has a
distribution, which is only on an average equal to the recoil
energy of the free case. This means that also the neutrinos have
the corresponding energy distribution and are therefore not
mono-energetic as in the 'free' decay.
\par%
We addressed this point already in the last sentence of our paper
\cite{GSI}: "Finally, an interesting case arises when the decaying
nucleus is not free but couples to the full phonon spectrum in the
lattice of a solid". Indeed, only for a true two-body EC-decay as,
for instance, from the ground state of a stored hydrogenic parent
ion to the ground state of its bare daughter ion without involving
a third object, a strict entanglement exists concerning momentum
and energy of the neutrino mass eigenstates on the one hand, and
of the corresponding recoiling nuclei on the other hand. We
discussed this point on p. 167, third paragraph, of \cite{GSI}.
\par%
Neither at Berkeley nor at GSI the generated neutrinos are
directly observed. In both experiments the time of the decay with
respect to the generation of the parent atom is precisely
determined, via the appearance of a characteristic K x-ray
(Berkeley), or via a sudden change of the mass of the stored ion
(GSI). We argue that only the latter case represents a true
two-body beta decay. Indeed, we get from the precisely determined
change of the mass the direct, time-resolved and complete
information at the hadronic vertex, i.e. on the transformation of
a proton to a neutron, as well as at the leptonic vertex, i.e. on
the annihilation of the K-shell electron and, thus, on the
generation of a neutrino in the electron-flavour eigenstate
(supposing lepton number conservation in the weak decay). This
knowledge could be the necessary condition for observing any kind
of interference in those decays.
\par%
It is interesting to note that the observed modulation frequency,
if indeed due to the interference of two neutrino mass
eigenstates, corresponds to a very small neutrino and, thus,
daughter recoil energy difference of about $8\cdot10^{-16}$~eV.
This is much smaller than typical phonon energies excited by the
recoiling daughter nuclei in an aluminum lattice which are in the
order of meV. Thus, the modulations could be washed out in a solid
environment. The very small energy difference measured in the GSI
experiment is in the order of that expected for the squared
neutrino mass difference of $10^{-4}$~eV$^2$ as pointed out in our
paper \cite{GSI}.
\par%
Concerning almost all other questions mentioned, we fully agree
with the authors: A measurement of the EC-decay of helium-like
$^{142}$Pm ions should reveal the (probably small) differences to
the EC-decay of hydrogen-like ions (such time-resolved
measurements are planned, but not yet performed). And, without
doubt, the outcome of the three-body $\beta^+$ decay of $^{142}$Pm
is crucial for the interpretation of the GSI data. The--not
simple--evaluation of this data is still in progress.%
\\%


\begin{thebibliography}{9}
\bibitem{LBNL}
P.A.~Vetter {\it et al.}, arXiv:0807.0649v1 [nucl-ex]\\
\bibitem{GSI}
Yu.A.~Litvinov {\it et al.}, Phys. Lett. B{\bf 664} (2008) 162.
\end{thebibliography}
\end{document}